**Employing Second-Order Circular Suprasegmental Hidden Markov Models to Enhance Speaker Identification Performance in Shouted Talking Environments**

Ismail Shahin

Electrical and Computer Engineering Department

University of Sharjah

P. O. Box  27272

Sharjah, United Arab Emirates

Tel: (971) 6 5050967

Fax: (971) 6 5050877

E-mail: ismail@sharjah.ac.ae

# Abstract

Speaker identification performance is almost perfect in neutral talking environments; however, the performance is deteriorated significantly in shouted talking environments. This work is devoted to proposing, implementing and evaluating new models called Second-Order Circular Suprasegmental Hidden Markov Models (CSPHMM2s) to alleviate the deteriorated performance in the shouted talking environments. These proposed models possess the characteristics of both Circular Suprasegmental Hidden Markov Models (CSPHMMs) and Second-Order Suprasegmental Hidden Markov Models (SPHMM2s). The results of this work show that CSPHMM2s outperform each of: First-Order Left-to-Right Suprasegmental Hidden Markov Models (LTRSPHMM1s), Second-Order Left-to-Right Suprasegmental Hidden Markov Models (LTRSPHMM2s) and First-Order Circular Suprasegmental Hidden Markov Models (CSPHMM1s) in the shouted talking environments. In such talking environments and using our collected speech database, average speaker identification performance based on LTRSPHMM1s, LTRSPHMM2s, CSPHMM1s and CSPHMM2s is 74.6%, 78.4%, 78.7% and 83.4%, respectively. Speaker identification performance obtained based on CSPHMM2s is close to that obtained based on subjective assessment by human listeners.

**Keywords:** first-order circular suprasegmental hidden Markov models; first-order left-to-right suprasegmental hidden Markov models; second-order circular suprasegmental hidden Markov models; second-order left-to-right suprasegmental hidden Markov models; shouted talking environments; speaker identification.

# 1. Introduction

Speaker recognition is the process of automatically recognizing who is speaking on the basis of individual information embedded in speech signals. Speaker recognition involves two applications: speaker identification and speaker verification (authentication). Speaker identification is the process of finding the identity of the unknown speaker by comparing his/her voice with voices of registered speakers in the database. The comparison results are measures of the similarity from which the maximal quality is chosen. Speaker identification can be used in criminal investigations to determine the suspected persons who generated the voice recorded at the scene of the crime. Speaker identification can also be used in civil cases or for the media. These cases include calls to radio stations, local or other government authorities, insurance companies, monitoring people by their voices and many other applications.

Speaker verification is the process of determining whether the speaker identity is who the person claims to be. In this type of speaker recognition, the voiceprint is compared with the speaker voice model registered in the speech data corpus that is required to be verified. The result of comparison is a measure of the similarity from which acceptance or rejection of the verified speaker follows. The applications of speaker verification include using the voice as a key to confirm the identity claim of a speaker. Such services include banking transactions using a telephone network, database access services, security control for confidential information areas, remote access to computers, tracking speakers in a conversation or broadcast and many other applications.

Speaker recognition is often classified into closed-set recognition and open-set recognition. The closed-set refers to the cases that the unknown voice must come from a set of known speakers, while the open-set refers to the cases that the unknown voice may come from unregistered speakers. Speaker recognition systems could also be divided according to the speech modalities: text-dependent (fixed-text) recognition and text-independent (free-text) recognition. In the text-dependent recognition, the text spoken by the speaker is known; however, in the text-independent recognition, the system should be able to identify the unknown speaker from any text.

## 2. Motivation and Literature Review

Speaker recognition systems perform extremely well in neutral talking environments [1-4]; however, such systems perform poorly in stressful talking environments [5-13]. Neutral talking environments are defined as the talking environments in which speech is generated assuming that speakers are not suffering from any stressful or emotional talking conditions. Stressful talking environments are defined as the talking environments that cause speakers to vary their generation of speech from neutral talking condition to other stressful talking conditions such as shouted, loud and fast.

In literature, there are many studies that focus on speech recognition and speaker recognition fields in stressful talking environments [5-13]. Specifically, these two fields are investigated by very few researchers in shouted talking environments. Therefore, the number of studies that focus on the two fields in such talking environments is limited [7-11]. Shouted talking environments are defined as the talking environments in which when speakers shout, their aim is to produce a very loud acoustic signal, either to increase its range of transmission or its ratio to background noise [8-11]. Speaker recognition systems in shouted talking environments can be used in criminal investigations to identify the suspected persons who uttered voice in a shouted talking condition and in the applications of talking condition recognition systems. Talking condition recognition systems can be used in: medical applications, telecommunications, law enforcement and military applications [12].

Chen studied talker-stress-induced intraword variability and an algorithm that compensates for the systematic changes observed based on hidden Markov models (HMMs) trained by speech tokens in different talking conditions [7]. In four of his earlier studies, Shahin focused on enhancing speaker identification performance in shouted talking environments based on each of Second-Order Hidden Markov Models (HMM2s) [8], Second-Order Circular Hidden Markov Models (CHMM2s) [9], Suprasegmental Hidden Markov Models (SPHMMs) [10] and gender-dependent approach using SPHMMs [11]. He achieved speaker identification performance in such talking environments of 59.0%, 72.0%, 75.0% and 79.2% based on HMM2s, CHMM2s, SPHMM2s and gender-dependent approach using SPHMMs, respectively [8-11].

This paper aims at proposing, implementing and testing new models to enhance text-dependent speaker identification performance in shouted talking environments. The new proposed models are called Second-Order Circular

Suprasegmental Hidden Markov Models (CSPHMM2s). This work is a continuation to the work of the four previous studies in [8-11]. Specifically, the main goal of this work is to further improve speaker identification performance in such talking environments based on a combination of each of: HMM2s, CHMM2s and SPHMMs. This combination is called CSPHMM2s. We believe that CSPHMM2s are superior models to each of: First-Order Left-to-Right Suprasegmental Hidden Markov Models (LTRSPHMM1s), Second-Order Left-to-Right Suprasegmental Hidden Markov Models (LTRSPHMM2s) and First-Order Circular Suprasegmental Hidden Markov Models (CSPHMM1s). This is because CSPHMM2s possess the combined characteristics of each of LTRSPHMM1s, LTRSPHMM2s and CSPHMM1s. In this work, speaker identification performance in each of the neutral and shouted talking environments based on CSPHMM2s is compared separately with that based on each of: LTRSPHMM1s, LTRSPHMM2s and CSPHMM1s.

The rest of the paper is organized as follows. The next section overviews the fundamentals of SPHMMs. Section 4 summarizes LTRSPHMM1s, LTRSPHMM2s and CSPHMM1s. The details of CSPHMM2s are discussed in Section 5. Section 6 describes the collected speech data corpus adopted for the experiments. Section 7 is committed to discussing speaker identification algorithm and the experiments based on each of LTRSPHMM1s, LTRSPHMM2s, CSPHMM1s and CSPHMM2s. Section 8 discusses the results obtained in this work. Concluding remarks are given in Section 9.

### 3. Fundamentals of Suprasegmental Hidden Markov Models

SPHMMs have been developed, used and tested by Shahin in the fields of: speaker recognition [10, 11, 14] and emotion recognition [15]. SPHMMs have demonstrated to be superior models over HMMs for speaker recognition in each of the shouted [10, 11] and emotional talking environments [14]. SPHMMs have the ability to condense several states of HMMs into a new state called suprasegmental state. Suprasegmental state has the capability to look at the observation sequence through a larger window. Such a state allows observations at rates appropriate for the situation of modeling. For example, prosodic information can not be detected at a rate that is used for acoustic modeling. Fundamental frequency, intensity and duration of speech signals are the main acoustic parameters that describe prosody [16]. Suprasegmental observations encompass information about the pitch of the speech signal, information about the intensity of the uttered utterance and information about the duration of the relevant segment. These three parameters in addition to the speaking style feature have been adopted and used in the current work. Prosodic features of a unit of speech are called suprasegmental features since they affect all the segments of the unit. Therefore, prosodic events at the levels of: phone, syllable, word and utterance are modeled using suprasegmental states; on the other hand, acoustic events are modeled using conventional states.

Prosodic and acoustic information can be combined and integrated within HMMs as given by the following formula [17],

$$log\ P\left(\lambda^v, \Psi^v | O\right) = (1-\alpha).\ log\ P\left(\lambda^v | O\right) + \alpha.\ log\ P\left(\Psi^v | O\right) \qquad (1)$$

where $\alpha$ is a weighting factor. When:

$$\begin{cases} 0.5 > \alpha > 0 & \text{biased towards acoustic model} \\ 1 > \alpha > 0.5 & \text{biased towards prosodic model} \\ \alpha = 0 & \text{biased completely towards acoustic model and} \\ & \text{no effect of prosodic model} \\ \alpha = 0.5 & \text{no biasing towards any model} \\ \alpha = 1 & \text{biased completely towards prosodic model and} \\ & \text{no impact of acoustic model} \end{cases}$$

$\lambda^v$: is the acoustic model of the $v$th speaker.

$\Psi^v$: is the suprasegmental model of the $v$th speaker.

$O$: is the observation vector or sequence of an utterance.

$P(\lambda^v | O)$: is the probability of the $v$th HMM speaker model given the observation vector $O$.

$P(\Psi^v | O)$: is the probability of the $v$th SPHMM speaker model given the observation vector $O$. The reader can obtain more details about suprasegmental hidden Markov models from the references: [10, 11, 14, 15].

## 4. Overview of: LTRSPHMM1s, LTRSPHMM2s and CSPHMM1s

### 4.1. First-order left-to-right suprasegmental hidden Markov models

First-Order Left-to-Right Suprasegmental Hidden Markov Models have been derived from acoustic First-Order Left-to-Right Hidden Markov Models (LTRHMM1s). LTRHMM1s have been adopted in many studies in the areas of: speech, speaker and emotion recognition in the last three decades because phonemes follow strictly the left to right sequence [18-20]. Fig. 1 shows an example of a basic structure of LTRSPHMM1s that has been derived from LTRHMM1s. This figure shows an example of six first-order acoustic hidden

Markov states ($q_1$, $q_2$, ...,$q_6$) with a left-to-right transition, $p_1$ is a first-order suprasegmental state consisting of $q_1$, $q_2$ and $q_3$, $p_2$ is a first-order suprasegmental state composing of $q_4$, $q_5$ and $q_6$. The suprasegmental states $p_1$ and $p_2$ are arranged in a left-to-right form. $p_3$ is a first-order suprasegmental state which is made up of $p_1$ and $p_2$. $a_{ij}$ is the transition probability between the $i$th and the $j$th acoustic hidden Markov states, while $b_{ij}$ is the transition probability between the $i$th and the $j$th suprasegmental states.

In LTRHMM1s, the state sequence is a first-order Markov chain where the stochastic process is expressed in a 2-D matrix of a priori transition probabilities ($a_{ij}$) between states $s_i$ and $s_j$ where $a_{ij}$ are given as:

$$a_{ij} = \text{Prob}(q_t = s_j | q_{t-1} = s_i) \qquad (2)$$

In these acoustic models, it is assumed that the state-transition probability at time $t+1$ depends only on the state of the Markov chain at time $t$. More information about acoustic first-order left-to-right hidden Markov models can be found in the references: [21, 22].

**4.2. Second-order left-to-right suprasegmental hidden Markov models**

Second-Order Left-to-Right Suprasegmental Hidden Markov Models have been obtained from acoustic Second-Order Left-to-Right Hidden Markov Models (LTRHMM2s). As an example of such models, the six first-order acoustic left-to-right hidden Markov states of Fig. 1 are replaced by six second-order acoustic hidden Markov states arranged in the left-to-right form. The suprasegmental second-order states $p_1$ and $p_2$ are arranged in the left-to-right form. The suprasegmental state $p_3$ in such models becomes a second-order suprasegmental state.

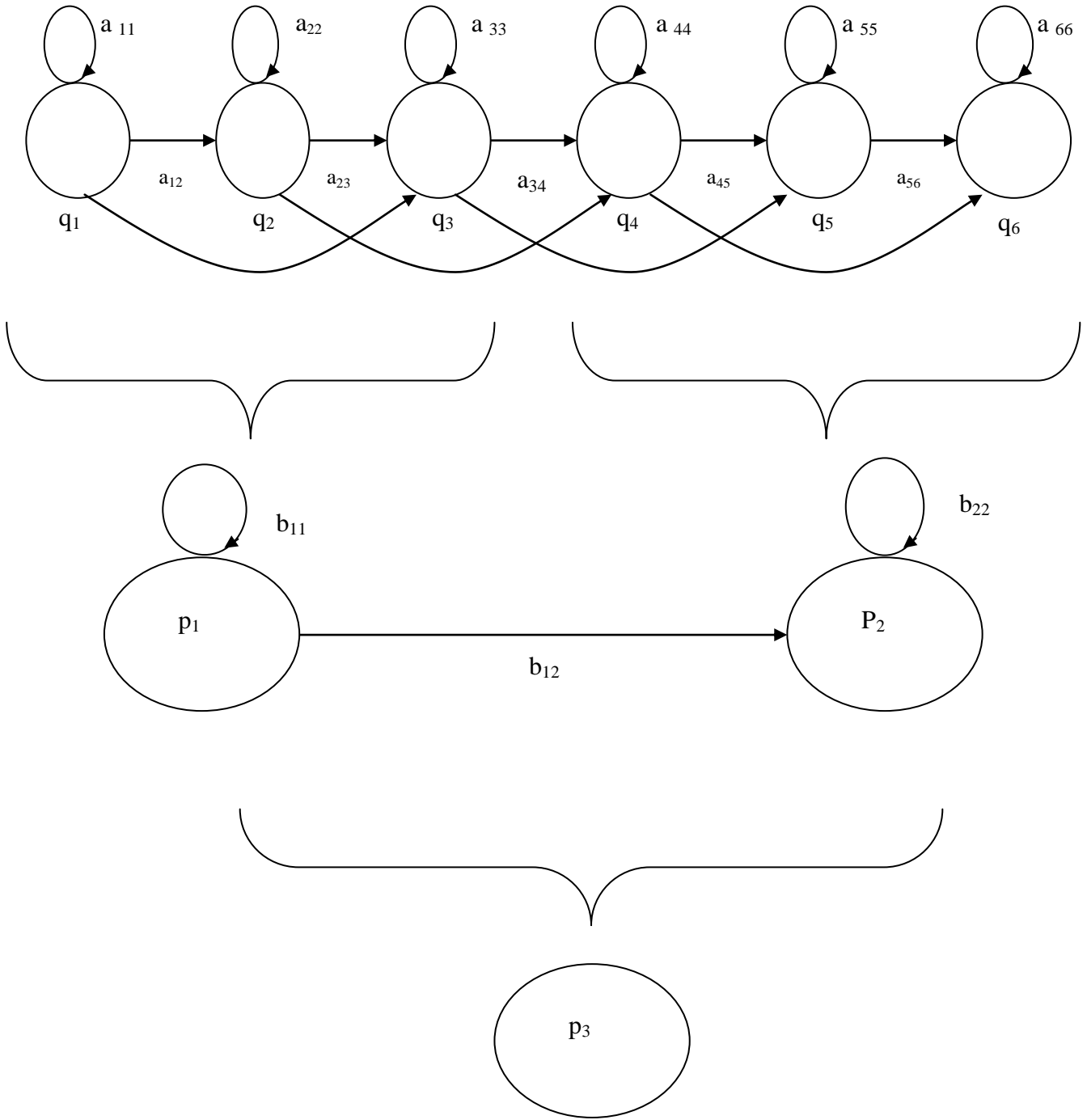

**Figure 1.** Basic structure of LTRSPHMM1s derived from LTRHMM1s

In LTRHMM2s, the state sequence is a second-order Markov chain where the stochastic process is specified by a 3-D matrix ($a_{ijk}$). Therefore, the transition probabilities in LTRHMM2s are given as [23]:

$$a_{ijk} = \text{Prob}(q_t = s_k | q_{t-1} = s_j, q_{t-2} = s_i) \qquad (3)$$

with the constraints,

$$\sum_{k=1}^{N} a_{ijk} = 1 \qquad N \geq i, j \geq 1$$

The state-transition probability in LTRHMM2s at time *t+1* depends on the states of the Markov chain at times *t* and *t-1*. The reader can find more information about acoustic second-order left-to-right hidden Markov models in the references: [8, 9, 23].

### 4.3. First-order circular suprasegmental hidden Markov models

First-Order Circular Suprasegmental Hidden Markov Models have been constructed from acoustic First-Order Circular Hidden Markov Models (CHMM1s). CHMM1s were proposed and used by Zheng and Yuan for speaker identification systems in neutral talking environments [24]. Shahin showed that these models outperform LTRHMM1s for speaker identification in shouted talking environments [9]. More details about CHMM1s can be obtained from the references: [9, 24].

Fig. 2 shows an example of a basic structure of CSPHMM1s that has been obtained from CHMM1s. This figure consists of six first-order acoustic hidden Markov states: $q_1, q_2, ..., q_6$ arranged in a circular form. $p_1$ is a first-order suprasegmental state consisting of $q_4$, $q_5$ and $q_6$. $p_2$ is a first-order suprasegmental state composing of $q_1$, $q_2$ and $q_3$. The suprasegmental states: $p_1$ and $p_2$ are arranged in a circular form. $p_3$ is a first-order suprasegmental state which is made up of $p_1$ and $p_2$.

### 5. Second-Order Circular Suprasegmental Hidden Markov Models

Second-Order Circular Suprasegmental Hidden Markov Models (CSPHMM2s) have been formed from acoustic Second-Order Circular Hidden Markov Models (CHMM2s). CHMM2s were proposed, used and examined by Shahin for speaker identification in each of the shouted and emotional talking environments [9, 14]. CHMM2s have shown to be superior models over each of LTRHMM1s, LTRHMM2s and CHMM1s because CHMM2s contain the characteristics of both CHMMs and HMM2s [9].

As an example of CSPHMM2s, the six first-order acoustic circular hidden Markov states of Fig. 2 are replaced by six second-order acoustic circular hidden Markov states arranged in the same form. $p_1$ and $p_2$ become second-order suprasegmental states arranged in a circular form. $p_3$ is a second-order suprasegmental state which is composed of $p_1$ and $p_2$.

Prosodic and acoustic information within CHMM2s can be merged into CSPHMM2s as given by the following formula,

$$log\ P\left(\lambda^v_{CHMM2s}, \Psi^v_{CSPHMM2s} \mid O\right) = (1-\alpha).\ log\ P\left(\lambda^v_{CHMM2s} \mid O\right)$$
$$+ \alpha.\ log\ P\left(\Psi^v_{CSPHMM2s} \mid O\right) \quad (4)$$

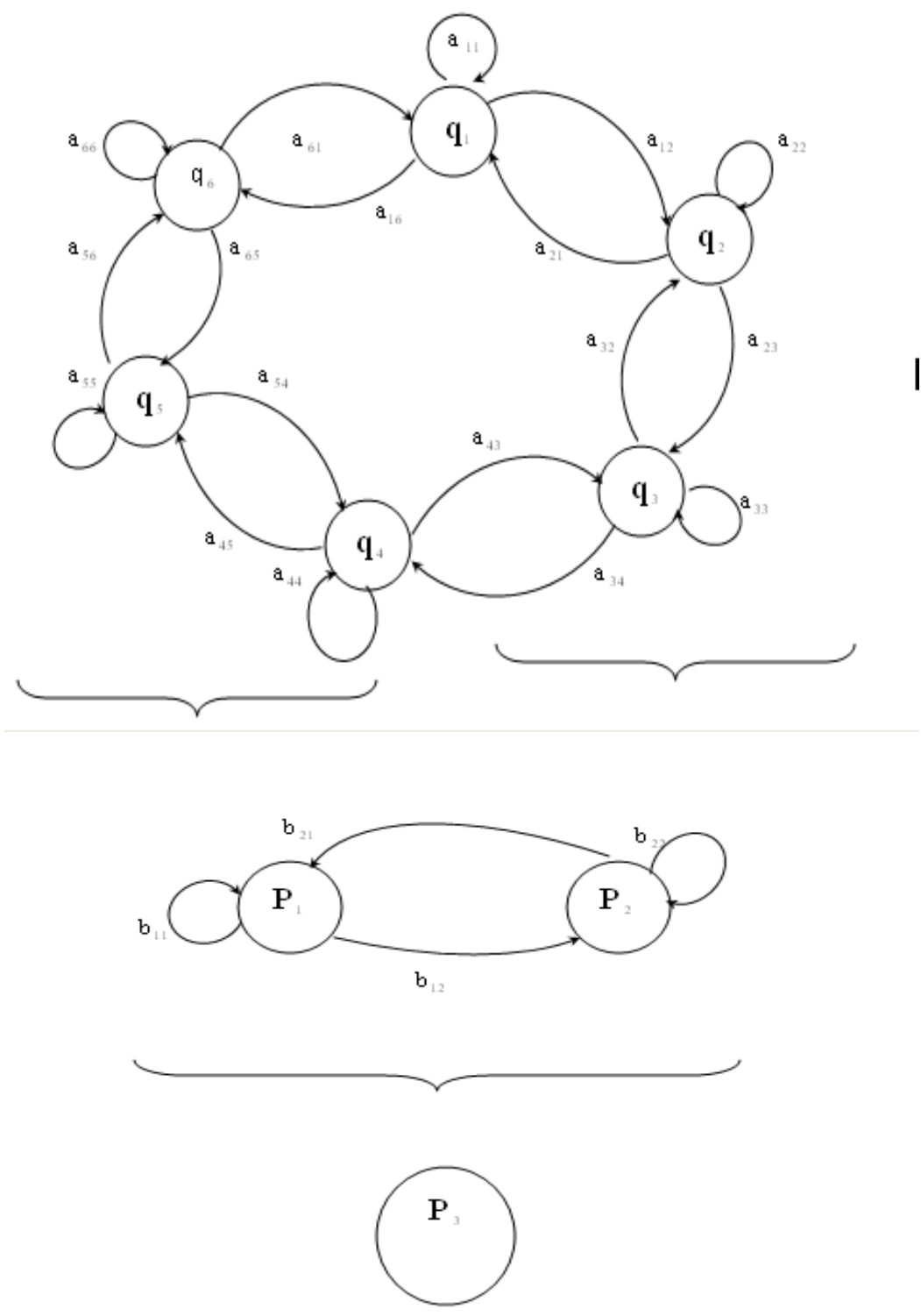

**Figure 2.** Basic structure of CSPHMM1s obtained from CHMM1s

where,

$\lambda^v_{CHMM2s}$ : is the acoustic second-order circular hidden Markov model of the *v*th speaker.

$\Psi^v_{CSPHMM2s}$ : is the suprasegmental second-order circular hidden Markov model of the *v*th speaker.

To the best of our knowledge, this is the first known investigation into CSPHMM2s evaluated for speaker identification in each of the neutral and shouted talking environments. CSPHMM2s are superior models over each of LTRSPHMM1s, LTRSPHMM2s and CSPHMM1s. This is because the characteristics of both CSPHMMs and SPHMM2s are combined and integrated into CSPHMM2s:

1. In SPHMM2s, the state sequence is a second-order suprasegmental chain where the stochastic process is specified by a 3-D matrix since the state-transition probability at time *t*+1 depends on the states of the suprasegmental chain at times *t* and *t*-1. On the other hand, the state sequence in SPHMM1s is a first-order suprasegmental chain where the stochastic process is specified by a 2-D matrix since the state-transition probability at time *t*+1 depends only on the suprasegmental state at time *t*. Therefore, the stochastic process that is specified by a 3-D matrix yields higher speaker identification performance than that specified by a 2-D matrix.

2. Suprasegmental chain in CSPHMMs is more powerful and more efficient than that possessed in LTRSPHMMs to model the changing statistical

characteristics that are available in the actual observations of speech signals.

## 6. Collected Speech Data Corpus

The proposed models in the current work have been evaluated using the collected speech data corpus. In this corpus, eight sentences were generated under each of the neutral and shouted talking conditions. These sentences were:

1) *He works five days a week.*
2) *The sun is shining.*
3) *The weather is fair.*
4) *The students study hard.*
5) *Assistant professors are looking for promotion.*
6) *University of Sharjah.*
7) *Electrical and Computer Engineering Department.*
8) *He has two sons and two daughters.*

Fifty (twenty five males and twenty five females) healthy adult native speakers of American English were asked to utter the eight sentences. The fifty speakers were untrained to avoid exaggerated expressions. Each speaker was separately asked to utter each sentence five times in one session (training session) and four times in another separate session (test session) under the neutral talking condition. Each speaker was also asked to generate each sentence nine times under the shouted talking condition for testing purposes. The total number of utterances in both sessions under both talking conditions was 7200.

The collected data corpus was captured by a speech acquisition board using a 16-bit linear coding A/D converter and sampled at a sampling rate of 16 kHz. The data corpus was a 16-bit per sample linear data. The speech signals were applied every 5 ms to a 30 ms Hamming window.

In this work, the features that have been adopted to represent the phonetic content of speech signals are called the Mel-Frequency Cepstral Coefficients (static MFCCs) and delta Mel-Frequency Cepstral Coefficients (delta MFCCs). These coefficients have been used in the stressful speech and speaker recognition fields because such coefficients outperform other features in the two fields and because they provide a high-level approximation of human auditory perception [25, 26, 27]. These spectral features have also been found to be useful in the classification of stress in speech [28, 29]. A 16-dimension feature analysis of both static MFCC and delta MFCC was used to form the observation vectors in each of LTRSPHMM1s, LTRSPHMM2s, CSPHMM1s and CSPHMM2s. The number of conventional states, $N$, was nine and the number of suprasegmental states was three (each suprasegmental state was composed of three conventional states) in each of LTRSPHMM1s, LTRSPHMM2s, CSPHMM1s and CSPHMM2s with a continuous mixture observation density was selected for each model.

## 7. Speaker Identification Algorithm Based on Each of LTRSPHMM1s, LTRSPHMM2s, CSPHMM1s and CSPHMM2s and the Experiments

The training session of LTRSPHMM1s, LTRSPHMM2s, CSPHMM1s and CSPHMM2s was very similar to the training session of the conventional LTRHMM1s, LTRHMM2s, CHMM1s and CHMM2s, respectively. In the training session of LTRSPHMM1s, LTRSPHMM2s, CSPHMM1s and CSPHMM2s (completely four separate training sessions), suprasegmental: first-order left-to-right, second-order left-to-right, first-order circular and second-order circular models were trained on top of acoustic: first-order left-to-right, second-order left-to-right, first-order circular and second-order circular models, respectively. For each model of this session, each speaker per sentence was represented by one reference model where each reference model was derived using five of the nine utterances per the same sentence per the same speaker under the neutral talking condition. The total number of utterances in each session was 2000.

In the test (identification) session for each of LTRSPHMM1s, LTRSPHMM2s, CSPHMM1s and CSPHMM2s (completely four separate test sessions), each one of the fifty speakers used separately four of the nine utterances per the same sentence (text-dependent) under the neutral talking condition. In another separate test session for each of LTRSPHMM1s, LTRSPHMM2s, CSPHMM1s and CSPHMM2s (completely four separate test sessions), each one of the fifty speakers used separately nine utterances per the same sentence under the shouted talking condition. The total number of utterances in each session was 5200. The probability of generating every utterance per speaker was separately computed based on each of LTRSPHMM1s, LTRSPHMM2s, CSPHMM1s and CSPHMM2s. For each one of these four suprasegmental models, the model with the highest probability was chosen as the output of speaker identification as given in the following formula per sentence per talking environment:

    a. In LTRSPHMM1s,

$$V^* = \arg \max_{50 \geq v \geq 1} \left\{ P\left(O \mid \lambda^v_{LTRHMM1s}, \Psi^v_{LTRSPHMM1s}\right) \right\} \quad (5)$$

where,

$O$: is the observation vector or sequence that belongs to the unknown speaker.

$\lambda^v_{LTRHMM1s}$: is the acoustic first-order left-to-right model of the $v$th speaker.

$\Psi^v_{LTRSPHMM1s}$: is the suprasegmental first-order left-to-right model of the $v$th speaker.

 b. In LTRSPHMM2s,

$$V^* = \arg \max_{50 \geq v \geq 1} \left\{ P\left(O \mid \lambda^v_{LTRHMM2s}, \Psi^v_{LTRSPHMM2s}\right) \right\} \quad (6)$$

where,

$\lambda^v_{LTRHMM2s}$: is the acoustic second-order left-to-right model of the $v$th speaker.

$\Psi^v_{LTRSPHMM2s}$: is the suprasegmental second-order left-to-right model of the $v$th speaker.

 c. In CSPHMM1s,

$$V^* = \arg \max_{50 \geq v \geq 1} \left\{ P\left(O \mid \lambda^v_{CHMM1s}, \Psi^v_{CSPHMM1s}\right) \right\} \quad (7)$$

where,

$\lambda^v_{CHMM1s}$: is the acoustic first-order circular model of the $v$th speaker.

$\Psi^v_{CSPHMM1s}$: is the suprasegmental first-order circular model of the $v$th speaker.

d. In CSPHMM2s,

$$V^* = \arg\max_{50 \geq v \geq 1} \left\{ P\left(O \mid \lambda^v_{CHMM2s}, \Psi^v_{CSPHMM2s}\right) \right\} \quad (8)$$

## 8. Results and Discussion

In the current work, CSPHMM2s have been proposed, implemented and evaluated for speaker identification systems in each of the neutral and shouted talking environments. To evaluate the proposed models, speaker identification performance based on such models is compared separately with that based on each of: LTRSPHMM1s, LTRSPHMM2s and CSPHMM1s in the two talking environments. In this work, the weighting factor ($\alpha$) has been chosen to be equal to 0.5 to avoid biasing towards any acoustic or prosodic model.

Table 1 shows speaker identification performance in each of the neutral and shouted talking environments using the collected database based on each of LTRSPHMM1s, LTRSPHMM2s, CSPHMM1s and CSPHMM2s. It is evident from this table that each of LTRSPHMM1s, LTRSPHMM2s, CSPHMM1s and CSPHMM2s perform almost perfect in the neutral talking environments. This is because each of the acoustic models: LTRHMM1s, LTRHMM2s, CHMM1s and CHMM2s yield high speaker identification performance in such talking environments as shown in Table 2.

Table 1

Speaker identification performance in each of the neutral and shouted talking environments using the collected database based on each of LTRSPHMM1s, LTRSPHMM2s, CSPHMM1s and CSPHMM2s

| Models | Gender | Speaker identification performance (%) | |
|---|---|---|---|
| | | Neutral talking environments | Shouted talking environments |
| LTRSPHMM1s | Male | 96.6 | 73.5 |
| | Female | 96.8 | 75.7 |
| | Average | 96.7 | 74.6 |
| LTRSPHMM2s | Male | 97.5 | 78.9 |
| | Female | 97.5 | 77.9 |
| | Average | 97.5 | 78.4 |
| CSPHMM1s | Male | 97.4 | 78.3 |
| | Female | 98.4 | 79.1 |
| | Average | 97.9 | 78.7 |
| CSPHMM2s | Male | 98.9 | 82.9 |
| | Female | 98.7 | 83.9 |
| | Average | 98.8 | 83.4 |

A statistical significance test has been performed to show whether speaker identification performance differences (speaker identification performance based on CSPHMM2s and that based on each of LTRSPHMM1s, LTRSPHMM2s and CSPHMM1s in each of the neutral and shouted talking environments) are real or simply due to statistical fluctuations. The statistical significance test has been carried out based on the Student $t$ Distribution test as given by the following formula,

Table 2

Speaker identification performance in each of the neutral and shouted talking environments using the collected database based on each of LTRHMM1s, LTRHMM2s, CHMM1s and CHMM2s

| Models | Gender | Speaker identification performance (%) | |
|---|---|---|---|
| | | Neutral talking environments | Shouted talking environments |
| LTRHMM1s | Male | 92.3 | 28.5 |
| | Female | 93.3 | 29.3 |
| | Average | 92.8 | 28.9 |
| LTRHMM2s | Male | 94.4 | 59.4 |
| | Female | 94.6 | 58.6 |
| | Average | 94.5 | 59.0 |
| CHMM1s | Male | 94.8 | 58.5 |
| | Female | 94.2 | 59.5 |
| | Average | 94.5 | 59.0 |
| CHMM2s | Male | 96.6 | 72.8 |
| | Female | 96.8 | 74.6 |
| | Average | 96.7 | 73.7 |

$$t_{model1, model2} = \frac{\overline{x}_{model1} - \overline{x}_{model2}}{SD_{pooled}} \quad (9)$$

where,

$\overline{x}_{model\ 1}$: is the mean of the first sample (model 1) of size *n*.

$\overline{x}_{model\ 2}$: is the mean of the second sample (model 2) of the same size.

$SD_{pooled}$: is the pooled standard deviation of the two samples (models) given as,

$$SD_{pooled} = \sqrt{\frac{SD_{model\ 1}^2 + SD_{model\ 2}^2}{n}} \quad (10)$$

where,

$SD_{model\ 1}$: is the standard deviation of the first sample (model 1) of size *n*.

SD $_{model\ 2}$: is the standard deviation of the second sample (model 2) of the same size.

In this work, the calculated *t* values in each of the neutral and shouted talking environments using the collected database between CSPHMM2s and each of LTRSPHMM1s, LTRSPHMM2s and CSPHMM1s are given in Table 3. In the neutral talking environments, each calculated *t* value is less than the tabulated critical value at *0.05* significant level $t_{0.05} = 1.645$. On the other hand, in the shouted talking environments, each calculated *t* value is greater than the tabulated critical value $t_{0.05} = 1.645$. Therefore, CSPHMM2s are superior models over each of LTRSPHMM1s, LTRSPHMM2s and CSPHMM1s in the shouted talking environments. This is because CSPHMM2s possess the combined characteristics of each of LTRSPHMM1s, LTRSPHMM2s and CSPHMM1s as was discussed in Section 5. This superiority becomes less in the neutral talking environments because the acoustic models: LTRHMM1s, LTRHMM2s and CHMM1s perform well in such talking environments as shown in Table 2.

In one of his previous studies, Shahin showed that CHMM2s contain the characteristics of each of LTRHMM1s, LTRHMM2s and CHMM1s. Therefore, the enhanced speaker identification performance based on CHMM2s is the resultant of speaker identification performance based on the combination of each of the three acoustic models as shown in Table 2. Since CSPHMM2s are derived from CHMM2s, the improved speaker identification performance in shouted talking environments based on CSPHMM2s is the resultant of the enhanced speaker identification performance based on each of the three suprasegmental models as shown in Table 1.

Table 3

The calculated *t* values in each of the neutral and shouted talking environments using the collected database between CSPHMM2s and each of LTRSPHMM1s, LTRSPHMM2s and CSPHMM1s

| | Calculated *t* value | |
|---|---|---|
| t <sub>model 1, model 2</sub> | Neutral environments | Shouted environments |
| t <sub>CSPHMM2s, LTRSPHMM1s</sub> | 1.231 | 1.874 |
| t <sub>CSPHMM2s, LTRSPHMM2s</sub> | 1.347 | 1.755 |
| t <sub>CSPHMM2s, CSPHMM1s</sub> | 1.452 | 1.701 |

Table 2 yields speaker identification performance in each of the neutral and shouted talking environments based on each of the acoustic models: LTRHMM1s, LTRHMM2s, CHMM1s and CHMM2s. Speaker identification performance achieved in this work in each of the neutral and shouted talking environments is consistent with that obtained in *Ref.* [9] using a different speech database (forty speakers uttering ten isolated words in each of the neutral and shouted talking environments) [9].

Table 4 gives the calculated *t* values between each suprasegmental model and its corresponding acoustic model in each of the neutral and shouted talking environments using the collected database. This table shows evidently that each suprasegmental model outperforms its corresponding acoustic model in each talking environment since each calculated *t* value in this table is greater than the tabulated critical value $t_{0.05} = 1.645$.

Table 4

The calculated *t* values between each suprasegmental model and its corresponding acoustic model in each of the neutral and shouted talking environments using the collected database

|  | Calculated *t* value | |
|---|---|---|
| t $_{\text{sup. model, acoustic model}}$ | Neutral environments | Shouted environments |
| t $_{\text{LTRSPHMM1s, LTRHMM1s}}$ | 1.677 | 1.785 |
| t $_{\text{LTRSPHMM2s, LTRHMM2s}}$ | 1.686 | 1.793 |
| t $_{\text{CSPHMM1s, CHMM1s}}$ | 1.697 | 1.887 |
| t $_{\text{CSPHMM2s, CHMM2s}}$ | 1.702 | 1.896 |

Four more experiments have been separately conducted in this work to evaluate the results achieved based on CSPHMM2s. The four experiments are:

1. The new proposed models have been tested using a well-known speech database called Speech Under Simulated and Actual Stress (SUSAS). SUSAS database was designed originally for speech recognition under neutral and stressful talking conditions [30]. In the present work, isolated words recorded at 8 kHz sampling rate were used under each of the neutral and angry talking conditions. Angry talking condition has been used as an alternative to the shouted talking condition since the shouted talking condition can not be entirely separated from the angry talking condition in our real life [8]. Thirty different utterances uttered by seven speakers (four males and three females) in each of the neutral and angry talking conditions have been chosen to assess the proposed models. This number of speakers is very limited compared to the number of speakers used in the collected speech database.

Fig. 3 illustrates speaker identification performance in each of the neutral and angry talking conditions using SUSAS database based on each of LTRSPHMM1s, LTRSPHMM2s, CSPHMM1s and CSPHMM2s. This figure shows apparently that speaker identification performance based on each model is almost ideal in the neutral talking condition. Based on each model, speaker identification performance using the collected database is very close to that using SUSAS database.

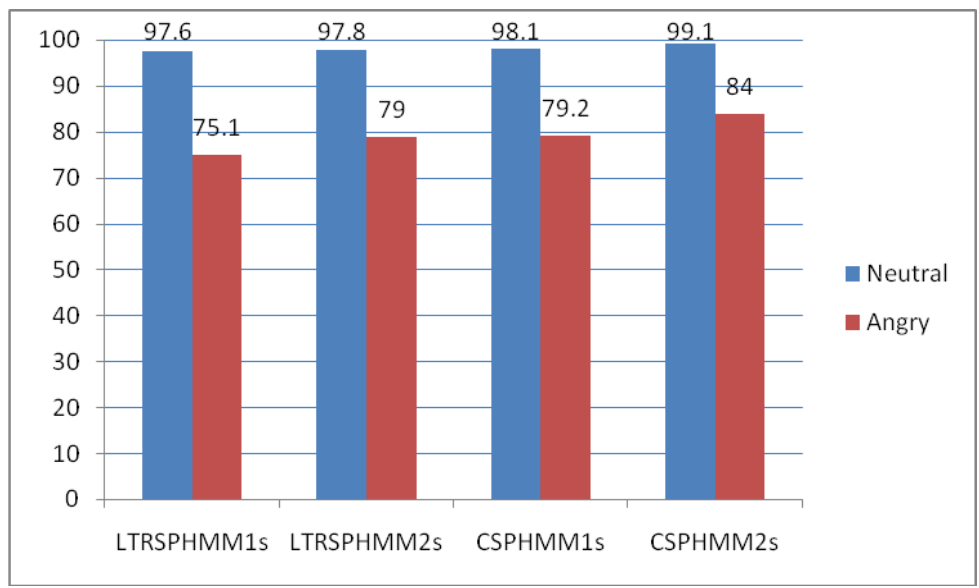

**Figure 3.** Speaker identification performance in each of the neutral and angry talking conditions using SUSAS database based on each of LTRSPHMM1s, LTRSPHMM2s, CSPHMM1s and CSPHMM2s

Table 5 yields the calculated $t$ values in each of the neutral and angry talking conditions using SUSAS database between CSPHMM2s and each of LTRSPHMM1s, LTRSPHMM2s and CSPHMM1s. This table demonstrates that CSPHMM2s lead each of LTRSPHMM1s, LTRSPHMM2s and CSPHMM1s in the angry talking condition.

Table 5

The calculated $t$ values in each of the neutral and angry talking conditions using SUSAS database between CSPHMM2s and each of LTRSPHMM1s, LTRSPHMM2s and CSPHMM1s

|  | Calculated $t$ value | |
| --- | --- | --- |
| $t_{\text{model 1, model 2}}$ | Neutral condition | Angry condition |
| $t_{\text{CSPHMM2s, LTRSPHMM1s}}$ | 1.345 | 1.783 |
| $t_{\text{CSPHMM2s, LTRSPHMM2s}}$ | 1.398 | 1.805 |
| $t_{\text{CSPHMM2s, CSPHMM1s}}$ | 1.499 | 1.795 |

Shahin reported in one of his previous studies speaker identification performance of 99.0% and 97.8% based on LTRSPHMM1s and gender-dependent approach using LTRSPHMM1s, respectively, in the neutral talking condition using SUSAS database [10, 11]. In the angry talking condition using the same database, Shahin achieved speaker identification performance of 79.0% and 79.2% based on LTRSPHMM1s and gender-dependent approach using LTRSPHMM1s, respectively [10, 11]. Based on using SUSAS database in each of the neutral and angry talking conditions, the results obtained in this experiment are consistent with those reported in some previous studies [10, 11].

2. The new proposed models have been tested for different values of the weighting factor ($\alpha$). Fig. 4 shows speaker identification performance in each of the neutral and shouted talking environments based on CSPHMM2s using the collected database for different values of $\alpha$ (0.0, 0.1, 0.2, ..., 0.9, 1.0). This figure indicates that increasing the value of $\alpha$ has a significant impact on enhancing speaker identification performance in the shouted talking environments. On the other hand, increasing the value of $\alpha$ has a less effect on improving the performance in the neutral talking environments. Therefore, suprasegmental hidden Markov models have more influence on speaker identification performance in the shouted talking environments than acoustic hidden Markov models.

3. A statistical cross-validation technique has been carried out to estimate the standard deviation of the recognition rates in each of the neutral and shouted talking environments based on each of LTRSPHMM1s, LTRSPHMM2s, CSPHMM1s and CSPHMM2s. Cross-validation technique has been performed separately for each model as follows: the entire collected database (7200 utterance per model) is partitioned at random into five subsets per model. Each subset is composed of 1440 utterance (400 utterance are used in the training session and the remaining are used in the evaluation session). Based on these five subsets per model, the standard deviation per model is calculated. The values are summarized in Fig. 5. Based on this figure, cross-validation technique shows that the calculated values of standard deviation are very low. Therefore, it is apparent that speaker identification performance in each of the neutral and shouted talking environments based on each of LTRSPHMM1s, LTRSPHMM2s, CSPHMM1s and CSPHMM2s using the five subsets per model is very close to that using the entire database (very slight fluctuations).

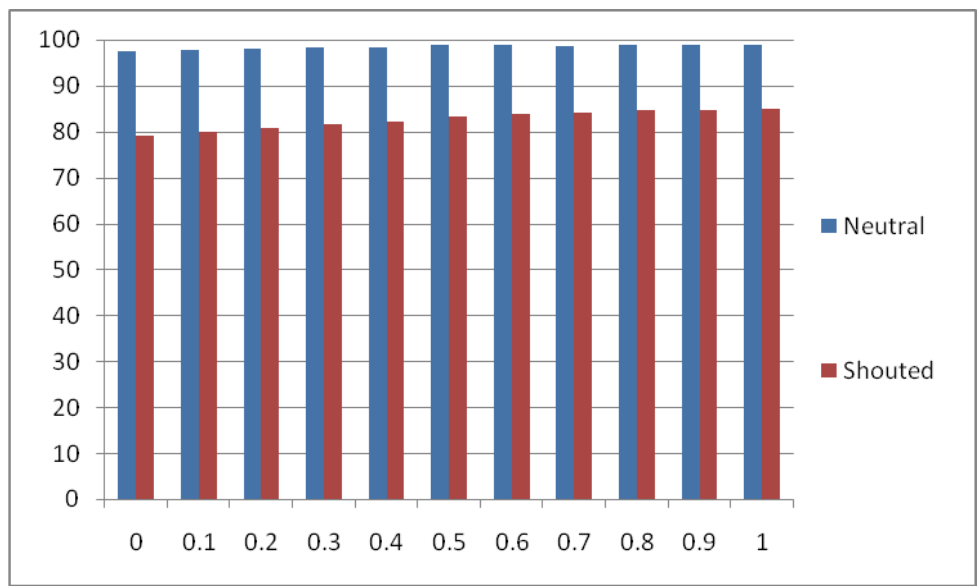

**Figure 4.** Speaker identification performance in each of the neutral and shouted talking environments based on CSPHMM2s using the collected database for different values of $\alpha$

4. An informal subjective assessment of the new proposed models using the collected speech database has been performed with ten nonprofessional listeners (human judges). A total of 800 utterance (fifty speakers, two talking environments and eight sentences) was used in this assessment. During the evaluation, each listener was asked to identify the unknown speaker in each of the neutral and shouted talking environments (completely two separate talking environments) for every test utterance. The average speaker identification performance in the neutral and shouted talking environments was 94.7% and 79.3%, respectively. These averages are very close to the achieved averages in the present work based on CSPHMM2s.

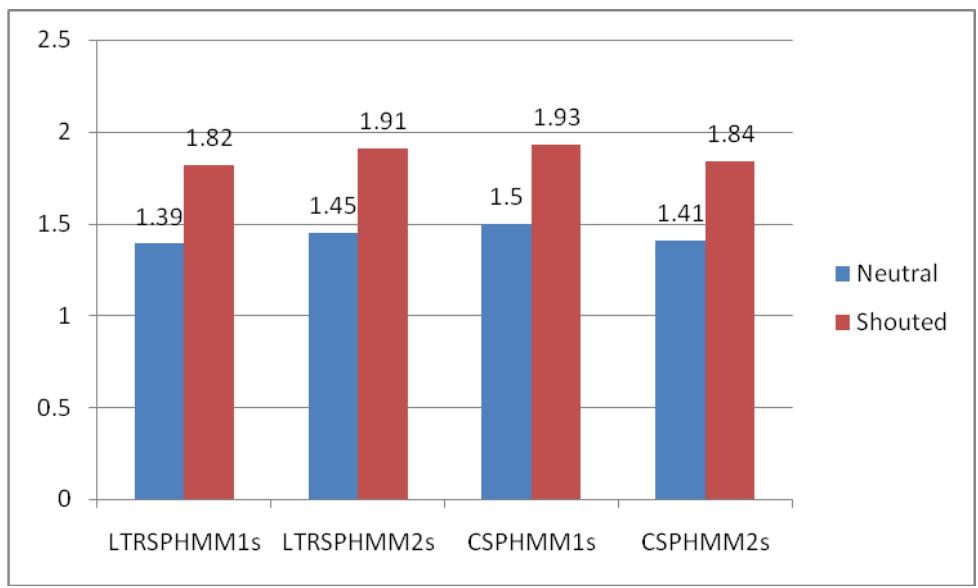

**Figure 5.** The calculated standard deviation values using statistical cross-validation technique in each of the neutral and shouted talking environments based on each of LTRSPHMM1s, LTRSPHMM2s, CSPHMM1s and CSPHMM2s

## 9. Concluding Remarks

In this work, CSPHMM2s have been proposed, implemented and evaluated to enhance speaker identification performance in the shouted/angry talking environments. Several experiments have been separately conducted in such talking environments using different databases based on the new proposed models. The current work shows that CSPHHM2s are superior models over each of: LTRSPHMM1s, LTRSPHMM2s and CSPHMM1s in each of the neutral and shouted/angry talking environments. This is because CSPHHM2s possess the combined characteristics of each of LTRSPHMM1s, LTRSPHMM2s and CSPHMM1s. This superiority is significant in the shouted/angry talking environments; however, it is less significant in the neutral talking environments. This is because the conventional HMMs perform extremely well in the neutral talking environments. Using CSPHMM2s for speaker identification systems increases nonlinearly the computational cost and the training requirements needed compared to using each of LTRSPHMM1s and CSPHMM1s for the same systems.

For future work, we plan to apply the proposed models to speaker identification systems in emotional talking environments. These models can also be applied to speaker verification systems in each of the shouted and emotional talking environments and to multi-language speaker identification systems.

## Acknowledgements

The author wishes to thank Prof. Mohammad Fraiwan Al-Saleh/ Prof. of Statistics at the University of Sharjah for his valuable help in the statistical part of this work.

## References

[1] K. R. Farrell, R. J. Mammone and K. T. Assaleh, "Speaker recognition using neural networks and conventional classifiers," IEEE Transaction on Speech and Audio Processing, Vol. 2, January 1994, pp. 194-205.

[2] K. Yu, J. Mason and J. Oglesby, "Speaker recognition using hidden Markov models, dynamic time warping and vector quantization," IEE Proceedings on Vision, Image and Signal Processing, Vol. 142, No. 5, October 1995, pp. 313-318.

[3] D. A. Reynolds, "Automatic speaker recognition using Gaussian mixture speaker models," The Lincoln Laboratory Journal, Vol. 8, No. 2, 1995, pp. 173-192.

[4] S. Furui, "Speaker-dependent-feature-extraction, recognition and processing techniques," Speech Communication, Vol. 10, March 1991, pp. 505-520.

[5] S. E. Bou-Ghazale and J. H. L. Hansen, "A comparative study of traditional and newly proposed features for recognition of speech under stress," IEEE Transaction on Speech and Audio Processing, Vol. 8, No. 4, July 2000, pp. 429-442.

[6] G. Zhou, J. H. L. Hansen and J. F. Kaiser, "Nonlinear feature based classification of speech under stress," IEEE Transaction on Speech and Audio Processing, Vol. 9, No. 3, March 2001, pp. 201-216.

[7] Y. Chen, "Cepstral domain talker stress compensation for robust speech recognition," IEEE Transaction on Acoustics, Speech, and Signal Processing, Vol. 36, No. 4, April 1988, pp. 433-439.

[8] I. Shahin, "Improving speaker identification performance under the shouted talking condition using the second-order hidden Markov models," EURASIP Journal on Applied Signal Processing, Vol. 5, issue 4, March 2005, pp. 482-486.

[9] I. Shahin, "Enhancing speaker identification performance under the shouted talking condition using second-order circular hidden Markov models" Speech Communication, Vol. 48, issue 8, August 2006, pp. 1047-1055.

[10] I. Shahin, "Speaker identification in the shouted environment using suprasegmental hidden Markov models," Signal Processing Journal, Vol. 88, issue 11, November 2008, pp. 2700-2708.

[11] I. Shahin, "Speaker identification in each of the neutral and shouted talking environments based on gender-dependent approach using SPHMMs," International Journal of Computers and Applications (In Press).

[12] J. H. L. Hansen, C. Swail, A. J. South, R. K. Moore, H. Steeneken, E. J. Cupples, T. Anderson, C. R. A. Vloeberghs, I. Trancoso and P. Verlinde, "The impact of speech under stress on military speech technology", NATO Research & Technology Organization RTO-TR-10, Vol. AC/323(IST)TP/ 5IST/TG-01, 2000.

[13] S. A. Patil and J. H. L. Hansen, "Detection of speech under physical stress: model development, sensor selection, and feature fusion," INTERSPEECH 2008, Brisbane, Australia, September 2008, pp. 817-820.

[14] I. Shahin, "Speaker identification in emotional environments," Iranian Journal of Electrical and Computer Engineering, Vol. 8, No. 1, Winter-Spring 2009, pp. 41-46.

[15] I. Shahin, "Speaking style authentication using suprasegmental hidden Markov models," University of Sharjah Journal of Pure and Applied Sciences, Vol. 5, No. 2, June 2008, pp. 41-65.

[16] J. Adell, A. Benafonte and D. Escudero, "Analysis of prosodic features: towards modeling of emotional and pragmatic attributes of speech," XXI Congreso de la Sociedad Española para el Procesamiento del Lenguaje Natural, SEPLN, Granada, Spain, September 2005.

[17] T. S. Polzin and A. H. Waibel, "Detecting emotions in Speech," Cooperative Multimodal Communication, Second International Conference 1998, CMC 1998.

[18] T. L. Nwe, S. W. Foo, L. C. De Silva, "Speech emotion recognition using hidden Markov models," Speech Communication, Vol. 41, issue 4, November 2003, pp. 603-623.

[19] D. Ververidis and C. Kotropoulos, "Emotional speech recognition: resources, features and methods," Speech Communication, Vol. 48, issue 9, September 2006, pp. 1162-1181.

[20] L. T. Bosch, "Emotions, speech and the ASR framework," Speech Communication, Vol. 40, issues 1-2, April 2003, pp. 213-225.

[21] L. R. Rabiner and B. H. Juang, Fundamentals of Speech Recognition, Prentice Hall, Eaglewood Cliffs, New Jersey, 1983.

[22] X. D. Huang, Y. Ariki and M. A. Jack, Hidden Markov Models for Speech Recognition, Edinburgh University Press, Great Britain, 1990.

[23] J. F. Mari, J. P. Haton and A. Kriouile, "Automatic word recognition based on second-order hidden Markov models," IEEE Transaction on Speech and Audio Processing, Vol. 5, No. 1, January 1997, pp. 22-25.

[24] C. Zheng and B. Z. Yuan, "Text-dependent speaker identification using circular hidden Markov models", IEEE International Conference on Acoustics, Speech and Signal Processing, S13.3, 1988, pp. 580-582.

[25] H. Bao, M. Xu and T. F. Zheng, "Emotion attribute projection for speaker recognition on emotional speech," INTERSPEECH 2007, Antwerp, Belgium, August 2007, pp. 758-761.

[26] A. B. Kandali, A. Routray and T. K. Basu, "Emotion recognition from Assamese speeches using MFCC features and GMM classifier," IEEE Region 10 Conference TENCON 2008, Hyderabad, India, November 2008, pp. 1-5.

[27] T. H. Falk and W. Y. Chan, "Modulation spectral features for robust far-field speaker identification," IEEE Transactions on Audio, Speech and Language Processing, Vol. 18, No. 1, January 2010, pp. 90-100.

[28] G. Zhou, J. H. L. Hansen and J. F. Kaiser, "Nonlinear feature based classification of speech under stress," IEEE Transaction on Speech and Audio Processing Journal, Vol. 9, No. 3, March 2001, pp. 201-216.

[29] J. H. L. Hansen and B. D. Womack, "Feature analysis and neural network-based classification of speech under stress," IEEE Transaction on Speech and Audio Processing Journal, Vol. 43, No. 4, July 1996, pp. 307-313.

[30] J.H.L. Hansen and S. Bou-Ghazale, "Getting started with SUSAS: A speech under simulated and actual stress database", EUROSPEECH-97: International Conference on Speech Communication and Technology, Rhodes, Greece, September 1997, Vol. 4, pp. 1743-1746.